\documentclass[aps,prc,twocolumn,showpacs,superscriptaddress,floatfix]{revtex4-1}  
\usepackage{graphicx}
\usepackage{amsmath}
\usepackage{multirow}
\newcommand {\nc} {\newcommand}
\nc {\beq} {\begin{eqnarray}} \nc {\eol} {\nonumber \\} \nc {\eeq} {\end{eqnarray}} \nc {\eeqn} [1] {\label{#1} \end{eqnarray}} \nc
{\eoln} [1] {\label{#1} \\} \nc {\ve} [1] {\mbox{\boldmath $#1$}} \nc {\rref} [1] {(\ref{#1})} \nc {\Eq} [1] {Eq.~(\ref{#1})} \nc
{\re} [1] {Ref.~\cite{#1}} \nc {\dem} {\mbox{$\frac{1}{2}$}} \nc {\arrow} [2] {\mbox{$\mathop{\rightarrow}\limits_{#1 \rightarrow
#2}$}} \nc {\la} {\langle} \nc {\ra} {\rangle}
\begin{document}
\title{Structure of the $^8$B and $^8$Li nuclei and the astrophysical $S_{17}(0)$-factor of the
$^7$Be($p,\gamma$)$^8$B direct capture process within a three-body model}
\author{E.M. Tursunov} \email{tursune@inp.uz} \affiliation {Institute of
Nuclear Physics, Academy of Sciences, 100214, Ulugbek, Tashkent, Uzbekistan}
\author{D.S. Toshova}
\email{toshova@inp.uz} \affiliation {Institute of Nuclear Physics,
Academy of Sciences, 100214, Ulugbek, Tashkent, Uzbekistan}
\author{S.A. Turakulov}
\email{turakulov@inp.uz} \affiliation {Institute of Nuclear Physics,
Academy of Sciences, 100214, Ulugbek, Tashkent, Uzbekistan}
\affiliation {Tashkent State Agrarian University, 100140 Tashkent,
Uzbekistan}

\begin{abstract}
The structure of the ground $(J^{\pi},T)=(2^+,1)$ and excited $(1^+,1)$ bound states of the $^8$B and $^8$Li nuclei is studied within
the framework of the $\alpha+^3$He($^3$H)+$p(n)$ three-body potential cluster model based on the hyperspherical Lagrange-mesh method.
The two-body $\alpha-^3$He($^3$H), $\alpha$-nucleon, and $^3$He($^3$H)-nucleon realistic potentials are taken from the literature
without any modification. Convergent theoretical estimates for the three-body binding energy and matter radius have been obtained
with the maximal hypermomentum $K_{max}=22$ and $K_{max}=28$ for the ground $2^+$ and excited $1^+$ bound states, respectively. For
the ground state of $^8$B, the extracted proton separation energy of 125.1 keV is slightly lower than the experimental data. The
asymptotic normalization coefficients (ANC) of the virtual transition $^8$B$\rightarrow^7$Be+$p$ are estimated self-consistently by
matching the overlap integral of the $^8$B three-body and the $^7$Be two-body wave functions with its asymptotics, which is expressed
by the Whittaker function. The obtained values are $C_{I=1}=0.211\pm 0.011$~fm$^{-1/2}$ and $C_{I=2}=0.739 \pm 0.008$~fm$^{-1/2}$ in
the spin 1 and spin 2 channels, respectively. For the ANC values of the $^8$Li$\rightarrow ^7$Li+$n$ virtual transition the estimates
$C_{I=1}=0.220\pm 0.008$~fm$^{-1/2}$ and $C_{I=2}=0.774\pm0.008$~fm$^{-1/2}$ are extracted. The ratio $C^2(^8 {\rm B})/C^2(^8 {\rm
Li})=0.912\pm 0.038$ implies a breaking of the mirror symmetry of the strong nuclear forces
due to the Coulomb interaction and the dynamical three-body effects [N.K. Timofeyuk et al., Phys.Rev. C78(2008), 054322].
 For the zero-energy astrophysical factor of the direct nuclear capture process
$^7$Be(p,$\gamma)^8$B an estimate $S_{17}(0)=22.492\pm0.503$ eV b was obtained based on the asymptotic theory developed by D. Baye
[Phys. Rev. C {\bf 62},065803 (2000)]. It was found that the most important contribution comes from the spin 2 channel with
$S^{(2)}_{17}(0)=20.838 \pm 0.472$ eV b, while the spin 1 channel yields $S^{(1)}_{17}(0)=1.654 \pm 0.173$ eV b. The theoretical
result for the astrophysical $S_{17}(0)$ factor is in a good agreement with the estimate
$S_{17}(0)=20.8\pm0.7{\rm(th)}\pm1.4{\rm(exp)}$  eV b of the Solar Fusion II, but larger than the recommended value
$S_{17}(0)=20.5\pm0.70$ eV b of the Solar Fusion III. At the same time, the obtained theoretical result is very close to the value of
22.4 eV b used in the most successful Solar Model BAR2M [W.~Yang and Z.~Tian, AJ {\bf 970} (2024), 38]. þ þ

\end{abstract}
 \maketitle

\section{Introduction}
\par The radiative capture reaction $^7$Be(p,$\gamma)^8$B plays a significant role in nuclear astrophysics, particularly in understanding
solar neutrino production processes within the proton-proton ($pp$)-chain~\cite{solar3,solar2}. The low-energy cross section of this
reaction cannot be measured directly with high precision due to the Coulomb barrier and the extremely small reaction probabilities at
stellar energies~\cite{solar2}. Therefore, different types of reliable theoretical techniques have been developed for the
extrapolation of the $S$-factor to zero energy ~\cite{baye00,tak18}. The estimation of the asymptotic normalization coefficient (ANC)
for the $a \rightarrow b + c$ virtual decay is a basis of the study of the direct capture processes in nuclear astrophysics
~\cite{mukh97,cas01}. The ANC plays the main role in reliably estimating of the cross-section (astrophysical $S$ factor) of important
reactions in the ultralow energy region, including zero energy~\cite{solar3}. Traditionally, the empirical ANC values are determined
from different types of nuclear transfer reactions ~\cite{azh99,tab06,olim16} within the distorted-wave Born approximation (DWBA).
The empirical value of the squared ANC $C^2=C^2_{p_{3/2}}+C^2_{p_{1/2}}= 0.465\pm0.041$ fm$^{-1}$ for the $^8{\rm
B}\rightarrow^7$Be$+p$ virtual decay has been extracted from the analysis of the experimental differential cross-sections of the
$^{10}$B($^7$Be,$^8$B)$^9$Be and $^{14}$N($^7$Be,$^8$B)$^{13}$C proton transfer reactions \cite{tab06}. On the other hand, the
analysis of the reaction $^7$Be$(d,n)^8$B predicted a value $0.613\pm0.060$ fm$^{-1}$ \cite{olim16} for this quantity. The ANC value
can also be calculated within purely realistic theoretical models such as the microscopic three-body cluster model \cite{desc04}, the
{\it ab initio} no-core shell model/resonating group method (NCSM/RGM) \cite{nav11}, and Skyrme Hartree-Fock theory \cite{chan03}, as
well as halo effective field theory (EFT)~\cite{zhang15,zhang18} at next-to-leading order. The three-body calculations of the ANC
value for the $^8{\rm B}\rightarrow^7$Be$+p$ virtual transition have been done in the hyperspherical harmonics method
\cite{grig98,grig99}. The latter method was limited by the value $K_{max}=12$ of the maximal hypermomentum, which requires careful
verification of the convergence of the results obtained there.

Particular attention should be paid to {\it ab initio} calculations performed within the framework of different approaches. Nollett
and Wiringa obtained a squared ANC value $C^2=C^2_{p_{3/2}}+C^2_{p_{1/2}}=0.537(27)$ ~fm$^{-1}$ using a combination of the realistic
Argonne $V_{18}$ two-nucleon and Urbana IX three-nucleon potentials within the variational Monte-Carlo method (VMC) ~\cite{nol11},
slightly different from the value of 0.509 fm$^{-1}$ of P. Navratil {\it et al.} within the no-core shell model/resonating group
method (NCSM/RGM) \cite{nav11}. The same NCSM was previously applied with a realistic CD-Bonn 2000 NN-potential yielding the result
$C^2$=0.587 ~fm$^{-1}$ for the squared ANC and $S_{17}(0)=22.1 \pm 1.0$ ev$\cdot$b for the zero-energy $S$ factor \cite{nav06}. The
new result of the same group within the NCSM with continuum based on the chiral effective field theory NN-potentials at the N$^4$LO
plus three-nucleon potentials turned out to be smaller with $S_{17}(0)=19.8 \pm 0.3$ ev$\cdot$b \cite{nav23}.
On the other hand, the experimental values of the $S$-wave $p+^7Be$ scattering lengths were not reproduced
in last work. Since the initial $S$-wave channel yields the main contribution to the synthesis process, its reliable description is
preferable.

In addition to above {\it ab initio} results, the estimates $C^2=C^2_{p_{3/2}}+C^2_{p_{1/2}}=0.564\pm0.023$ ~fm$^{-1}$ from the halo
effective field theory (EFT)~\cite{zhang15,zhang18} and $C^2=0.49\pm0.01$ fm$^{-1}$ obtained within the Bayesian approximation
~\cite{surer22} demonstrate clearly that the problem of determination of the ANC and the $S_{17}(0)$ astrophysical factor with a high
precision from the theory is still far from being resolved, since the range of ANC estimates from 0.480 fm$^{-1}$ to 0.587 ~fm$^{-1}$
is quite large. In addition, the zero-energy astrophysical $S$ factor was extracted by an indirect method \cite{tak18} by employing
the measured solar $^8$B and $^7$Be neutrino fluxes of the Borexino Collaboration \cite{agos19}, which yields $ S_{17}(0) \approx
19.5 \pm 1.9$ eV b.

One should note that the new SFIII recommended value $S_{17}(0)=20.5\pm0.70$ eV b \cite{solar3} mostly based on the {\it ab initio}
results of Ref. \cite{nav23}, is quite smaller than the SFII estimate $S_{17}(0)=20.8\pm0.7{\rm(th)}\pm1.4{\rm(exp)}$ eV b
\cite{solar2}. On the other hand, the most successful new Solar Model BAR2M \cite{solarmod} is based on the value 22.4 eV b for this
quantity, as well as previous Solar Models \cite{bach92}.

Another important point is that the potential models which account for cluster effects in nuclei explicitly, can complement the {\it
ab initio} methods in describing the astrophysical reactions with light nuclei. Recent studies of the  direct $\alpha(d,\gamma)^6$Li
capture process within the three-body model \cite{tur16,bt18,tur18,tur20,tb26} on the basis of the hyperspherical Lagrange-mesh
method have demonstrated that it can explain in details the data of direct measurements of the LUNA collaboration \cite{LUNA1,LUNA2}.
The three-body model was able to reproduce not only the absolute values, but also the energy and temperature dependence, of the
astrophysical $S$ factor and the reaction rates, respectively. The obtained result for the primordial abundance of the $^6$Li element
in the Big Bang nucleosynthesis (BBN) model is close to the Standard Model estimate \cite{tur18}. The most important results of that
research was a consistent realistic description of the isospin-forbidden $E1$ astrophysical S-factor due-to very small isotriplet
component of the $^6$Li nucleus of order 0.5$\%$ which yields the main contribution to the capture process at low energies below 100
keV.

A two-body effective potential model within a single-channel approximation \cite{tur21} provides a good description of the direct
astrophysical capture $^7$Be(p,$\gamma)^8$B process mostly due-to the correct reproduction of the $S$-wave $p+^7\rm Be$ scattering
length $a_{01}=17.34^{+1.11}_{-1.33}$ fm ~\cite{pan19}. The model reproduces the existing experimental data for the astrophysical S
factor in the energy region up to 6 MeV, while yielding the squared ANC value of $C^2=0.538^{+0.052}_{-0.050}$~fm$^{-1}$ and the
zero-energy factor $ S_{17}(0) \approx 20.51^{+2.02}_{-1.85}$ eV b, consistent with the Solar Fusion II (SFII) estimate
$S_{17}(0)=20.8\pm0.7{\rm(th)}\pm1.4{\rm(exp)}$  eV b \cite{solar2}. However, the model does not take into account the effects of the
$^7$Be core excitation, and the obtained results should be considered as a first approximation within the framework of the approach
based on the cluster model.

A determination of the ANC value for the $^8{\rm Li}\rightarrow ^7$Li$+n$ virtual decay is  important for the study of the
$^7$Li(n,$\gamma)^8$Li capture reaction in neutron-rich stars \cite{heil98, iga04}. This reaction is also useful for the study of the
mirror symmetry properties of strong nuclear forces. The most important value for the squared ANC value for the $^8{\rm
Li}\rightarrow^7$Li$+n$ virtual decay $C^2_{p_{3/2}}+C^2_{p_{1/2}}=0.432 \pm 0.044$~fm$^{-1}$ \cite{tra03} was obtained from the
analysis of the experimental differential cross-sections of the $^{13}$C($^7$Li,$^8$Li)$^{12}$C transfer reaction in the frame of the
DWBA. In particular, the authors of Refs.~\cite{shul96, grig98} have studied the three-body structure of the $^8$Li nucleus with
further application to the radiative capture process $^7$Li(n,$\gamma)^8$Li within the hyperspherical harmonics method. Burkova {\it
et al.}~\cite{bur21} have calculated the total cross sections of the neutron radiative capture reaction on $^7$Li at the low energy
region in the frame of the two-body potential cluster model.

The aim of the present work is to calculate the ANC values of the virtual decays $^8{\rm B}\rightarrow^7$Be$+p$ and $^8{\rm
Li}\rightarrow ^7$Li$+n$ in the spin 1 and spin 2 channels within a self-consistent three-body potential cluster model using
realistic $\alpha-p(n)$, $\alpha-^3$He($^3$H), and $p(n)-^3$He($^3$H) two-body interaction potentials. Furthermore, the obtained ANC
values will be applied for the estimation of the astrophysical factor of the radiative capture $^7$Be(p,$\gamma)^8$B reaction at the
zero-energy. For this purpose we calculate the characteristics of the $^8$B and $^8$Li nuclei in the framework of the
$\alpha+^3$He($^3$H)+$N$ three-body potential cluster model in the hyperspherical Lagrange-mesh method. The formalism of the
hyperspherical method is well known as presented in Refs. \cite{zhuk93,lin95,desc03}, and is widely used to study the three-body
structure, energy spectra, and spectroscopic information of light nuclei.

The article is organized as follows. The theoretical model will be briefly described in Section II, the two-body interaction
potentials are given in Section III, the numerical results are presented in Section IV, and the conclusions are drawn in the last
section.


\section{Theoretical model}

\begin{figure}[htbp]
\includegraphics[width=0.45\textwidth]{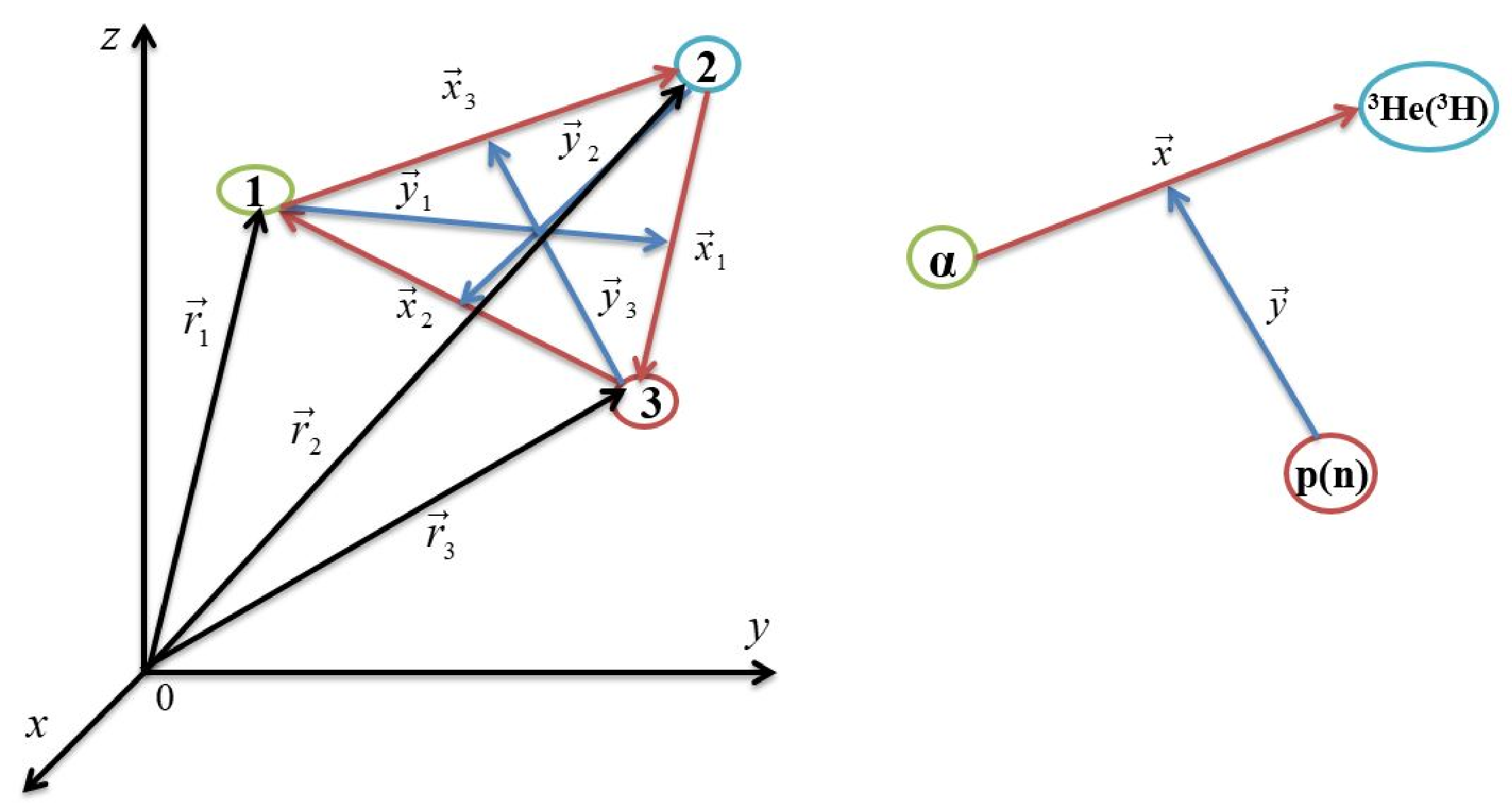} \caption{The Jacobi coordinates
for the three-body system $\alpha+^3{\rm He}(^3{\rm H})+{\rm p}({\rm
n})$.}\label{fig1}
\end{figure}

In this section, a theoretical model for the three-particle quantum system in hyperspherical coordinates is briefly described. The
mass numbers are given by $A_i$ (in units of the nucleon mass $m_N$), and the space coordinates are $\ve{r}_i$. The three-body
Hamiltonian is written as \cite{desc03,desc06,dan91}
\begin{eqnarray}
H=\sum_{i=1}^3T_i \ + \ \sum_{i>j=1}^3V_{ij}(\ve{r}_j-\ve{r}_i),
\label{eq1}
\end{eqnarray}
where $T_i$ is the kinetic energy of the $i-$th nucleus, and $V_{ij}$ is a nucleus-nucleus interaction potential. Starting from
coordinates $\ve{r}_i$ in Fig.~\ref{fig1}, we define the Jacobi coordinates $\ve{x}_k$ and $\ve{y}_k$ ($k=1,2,3$) as \cite{desc03}
\begin{multline}
\ve{x}_k = \sqrt{\mu_{ij}}(\ve{r}_j-\ve{r}_i), \\ \ve{y}_k =
\sqrt{\mu_{(ij)k}}
\left(\ve{r}_k-\frac{A_i\ve{r}_i+A_j\ve{r}_j}{A_i+A_j} \right),
\end{multline}
where $(i,j,k)$ is an even permutation of $(1,2,3)$ and the reduced
mass numbers are defined as
\begin{eqnarray}
 \mu_{ij}=A_i A_j/(A_i+A_j), \,\,\, \mu_{(ij)k} = \frac{(A_i+A_j) A_k} {A_i+A_j+A_k}.
\end{eqnarray}
The hyperradius $\rho$ and hyperangle $\alpha_k$ are then defined as $\rho^2 = x_k^2+y_k^2$,  $\alpha_k = \arctan{\frac{y_k}{x_k}}$,
respectively. The hyperangle $\alpha_k$ and the orientations $\Omega_{\ve{x}}$ and $\Omega_{\ve{y}}$ provide a set of angles
$\Omega_{5k}$. In this notation the kinetic energy reads \cite{desc03,desc06}
\begin{multline} T_{\rho}=\sum_{i=1}^3 T_i-T_{cm} \\=
-\frac{\hbar^2}{2m_N} \left( \frac{\partial^2}{\partial \rho^2} +
\frac{5}{\rho} \frac{\partial}{\partial \rho} -
\frac{K^2(\Omega_{5k})}{\rho^2}\right), \label{eq2}
\end{multline}
where $T_{cm}$ is the c.m. kinetic energy, and $K^2$ is a
five-dimensional angular momentum operator whose eigenfunctions
(with eigenvalues $K(K+4)$) are given by \cite{desc03,ray70}
\begin{eqnarray}
{\cal Y}^{\ell_x \ell_y}_{KLM_L}(\Omega_{5})&=&\phi^{\ell_x
\ell_y}_{K}(\alpha) \left[ Y_{\ell_x}(\Omega_{x})\otimes
Y_{\ell_y}(\Omega_{y}) \right] ^{LM_L}, \label{eq3}
\end{eqnarray}
\begin{multline}
\phi^{\ell_x \ell_y}_{K}(\alpha)= {\cal N}_K^{\ell_x \ell_y} (\cos
\alpha)^{\ell_x} (\sin \alpha)^{\ell_y}\\
\times
P_n^{(\ell_y+\mbox{$\frac{1}{2}$},\ell_x+\mbox{$\frac{1}{2}$})}(\cos
2\alpha), \label{eq4}
\end{multline}
where   ${\cal N}_K^{\ell_x \ell_y}$ is a normalization factor \cite{lin95},  $K$ is the hypermomentum,  $\ell_x$ and $\ell_y$ are
the orbital momenta associated with the coordinates $\ve{x}$ and $\ve{y}$, respectively, $n$ is a positive integer defined by
$n=(K-\ell_x-\ell_y)/2$, $P_n^{(\alpha,\beta)}(x)$ is the Jacobi polynomial. Introducing the spin component as a tensor product of
the proton (neutron) and $^3$He($^3$H) spin functions
$$\chi^{SM_S}(\xi_2,\xi_3)=[\chi^{1/2}(\xi_2)\otimes
\chi^{1/2}(\xi_3)]^{SM_S},$$ for the hyperspherical function with total momentum $J$ and its projection $M$, one obtaines: $$ {\cal
Y}^{JM}_{\gamma K}(\Omega_5,\xi_2,\xi_3)=\left[ {\cal Y}^{\ell_x \ell_y}_{KL}(\Omega_5) \otimes \chi^{S}(\xi_2,\xi_3) \right]
^{JM},$$ where the index $\gamma$ stands for ($\ell_x,\ell_y,L,S$).

The wave function $\Psi^{JM\pi}$, which is a solution of the Schr\"odinger equation associated with the Hamiltonian Eq.~(\ref{eq1}),
is expanded over basis functions as \cite{desc06}
\begin{multline}
\Psi^{JM\pi}(\rho,\Omega_5,\xi_2,\xi_3)=\rho^{-5/2}\sum_{\gamma K} {\chi}^{J\pi}_{\gamma K}(\rho) \\ \times{\cal Y}^{JM}_{\gamma
K}(\Omega_5,\xi_2,\xi_3), \label{eq5}
\end{multline}
where ${\chi}^{J\pi}_{\gamma K}(\rho)$ are hyperradial wave functions to be determined. Rigorously, the summation over ($\gamma K$)
involves an infinite number of terms. In practice, this expansion is limited by a maximal $K$ value, denoted as $K_{max}$. For weakly
bound states, it is well known that the convergence is rather slow, and therefore the large values of $K_{max}$ are required. The
radial functions ${\chi}^{J\pi}_{\gamma K}(\rho)$ are derived from a set of coupled differential equations \cite{desc06}
\begin{multline}
\left[-\frac{\hbar^2}{2m_N} \left( \frac{d^2}{d\rho^2} - \frac{{\cal
L}_K ({\cal L}_K + 1)}{\rho^2}\right) -E \right]
{\chi}^{J\pi}_{\gamma K}(\rho) \\+ \sum_{K' \gamma'} V^{J\pi}_{K
\gamma,K' \gamma'}(\rho)\, {\chi}^{J\pi}_{\gamma' K'}(\rho)=0,
\label{eq6}
\end{multline}
with ${\cal L}_K =K+3/2.$ The potential terms are given by the
contribution of the three nucleus-nucleus interactions
\begin{eqnarray}
V^{J\pi}_{K  \gamma,K' \gamma'}(\rho)=\sum_{i=1}^3 (V^{J\pi(Ni)}_{K
\gamma,K' \gamma'}(\rho)+V^{J\pi(Ci)}_{K  \gamma,K' \gamma'}(\rho)),
\label{eq7}
\end{eqnarray}
where the nuclear $(N)$ and Coulomb ($C)$ terms have been explicitly written. The system of coupled equations Eq.(\ref{eq6}) is
solved by expanding the trial function over the Lagrange-mesh basis function. The Pauli forbidden states in the three-body system are
eliminated with the help of the orthogonalizing pseudopotentials (OPP) method \cite{desc06}.

\section{Cluster-cluster interaction potentials}

\subsection{$\alpha-N$ interaction potential}
For the $\alpha-N$ interaction, a nuclear potential with odd-even splitting proposed by Voronchev {\it et al.} \cite{vor95} was used.
The potential includes a deep Pauli-forbidden state in the S-wave and provides an accurate description of the phase shifts for both
$\alpha-n$ and $\alpha-p$ elastic scattering in the $S-, P-, D-$ partial waves within the energy range of 0-20 MeV:

\begin{eqnarray}
V_{N-\alpha}(r)=V_{c}(r)+(\vec{l}\cdot \vec{s})V_{l s}(r)+V_{Coul}(r).    \label{eq8}
\end{eqnarray}

The radial dependencies of the central $V_c(r)$ and spin-orbital $V_{ls}(r)$ terms have been accurately parameterized using the
Gaussian functions
\begin{eqnarray}
V_{k}(r)=V_k \exp(-\eta_k^2 r^2), \,\, k=c, ls \label{eq9}
\end{eqnarray}
with the parameters given in Table \ref{table1}. The Coulomb part of the $p-\alpha$ interaction $V_{Coul}(r)= 2e^2erf(0.83 R)/R $
\cite{rei70}.

\begin{table}[htbp]
\centering \caption{The potential parameters of the $\alpha-N$
interaction \cite{vor95}.}\label{table1}
\begin{tabular}{cccccccc}
\hline\hline Parity & $V_c$, MeV & $V_{l s}$, MeV & $\eta_c$, fm$^{-1}$ &$\eta_{l s}$, fm$^{-1}$ \\
\hline Even & -66.580 & -12.169 & 0.6203 & 0.8032  \\
Odd &-46.303 & -15.931 & 0.4321 & 0.6282   \\ \hline\hline
\end{tabular}
\end{table}

\subsection{$\alpha-^3$He($^3$H) interaction potential}
The central $\alpha-^3$He and $\alpha-^3$H two-body potentials are also taken in a simple Gaussian form \cite{dub10}:
 \begin{eqnarray}
 V(r)=V_0 \exp(-\alpha_0 r^2)+V_{C}(r),
\label{eq10}
 \end{eqnarray}
 where the Coulomb part is given as
\begin{eqnarray}
 V_{C}(r)=
\left\{
\begin{array}{lc}
Z_1 Z_2 e^2/r &  {\rm if} \,\, r>R_c, \\
Z_1 Z_2 e^2 \left(3-{r^2}/{R_c^2}\right)/(2R_c) & {\rm if} \,\,\leq
R_c,
\end{array}
\right. \label{eq11}
\end{eqnarray}

with the Coulomb radius $R_c=3.095$ fm, and charge numbers $Z_1$, $Z_2$ of the first and second clusters, respectively. The
parameters $V_0$ of the central part of the potential are specified for each partial wave (see Table \ref{table2}), while a value of
$\alpha_0=0.15747$ fm$^{-2}$ is the same for all partial waves \cite{dub10}. The potential had been constructed according to the
classification of orbital states of the Young schemes. It contains two Pauli forbidden states in the $S$-waves, and a single
forbidden state in each of the partial $P-$ and $D-$waves. The energy values of these states are presented in the second row of the
table.

\begin{table}[htbp]
\caption {Values of the depth parameter $V_0$ of the $\alpha - ^3${\rm He} ($^3$H) potential for different partial waves
\cite{dub10}.} \label{table2}
\begin{tabular}{cccccccc}
\hline\hline $^{2s+1}L_J$ & $^2S_{1/2}$ & $^2P_{1/2}$ & $^2P_{3/2}$
& $^2D_{3/2}$ & $^2D_{5/2}$ & $^2F_{5/2}$ & $^2F_{7/2}$\\ \hline
$V_0$, MeV  & -67.47 &  -81.92 &-83.83 & -66.0 & -69.0 &-75.9&-84.8
\\
E$_{\rm {FS}}$, MeV & -36.0; -7.4 & -27.5 &-28.4 &-2.9&-4.1&-&-\\
\hline\hline
\end{tabular}
\end{table}
The interaction potential describes the experimental data \cite{boykin,hardy,spiger} quite accurately for the $^3$He$+\alpha$ and
$^3$H$+\alpha$ phase shifts in the $S-, P-, D-$ and $F-$ partial waves.

\subsection{$p(n)-^3$He($^3$H) interaction potential}
The $p-^3$He and $n-^3$H two-body potentials are taken as a sum of the Gaussian attractive and the exponential repulsive nuclear
parts \cite{dub97} with an additional Coulomb term:
 \begin{eqnarray}
 V(r)=V_0 \exp(-\alpha_0 r^2)+V_1 \exp(-\beta_0 r) +V_{C}(r). \label{eq12}
 \end{eqnarray}
The Coulomb part $V_{C}(r)$ of the potential is given with the point-like charge distribution. It also contains a Pauli forbidden
state in the $S$-wave.  The potential parameters are given in Table \ref{table3}. The last column contains energies of the
corresponding forbidden states in the singlet and triplet S-waves.

\begin{table}[htbp]
\caption {Parameters of the $p(n)-^3$He($^3$H) interaction potential in different partial waves \cite{dub97}.} \label{table3}
\begin{tabular}{cccccc}
\hline\hline
 $^{2s+1}L_J$ & $V_0$, MeV & $\alpha_0$, fm$^{-2}$ & $V_1$, MeV & $\beta_0$, fm$^{-1}$  & E$_{\rm {FS}}$, MeV \\ \hline
 $S=0$ &  & & & & \\ \hline
 even & -110.0 & 0.37 & 45.0 & 0.67 & -9.0(-11.4) \\
 odd  & -14.0 & 0.10 &  &  &  \\ \hline
 $S=1$ &  & & & & \\ \hline
 even & -43.0 & 0.26 & & &-3.6(-5.3) \\
 $^3P_0$ & -10.0 & 0.10 & & &- \\
 $^3P_1$ & -15.0 & 0.10 & & &- \\
 $^3P_2$ & -20.0 & 0.10 & & &- \\ \hline\hline

\end{tabular}
\end{table}

\section{Numerical results}
\subsection{Energies of the ground 2$^+$ and first excited 1$^+$ bound states.}

\begin{figure}[htbp]
\includegraphics[width=0.45\textwidth]{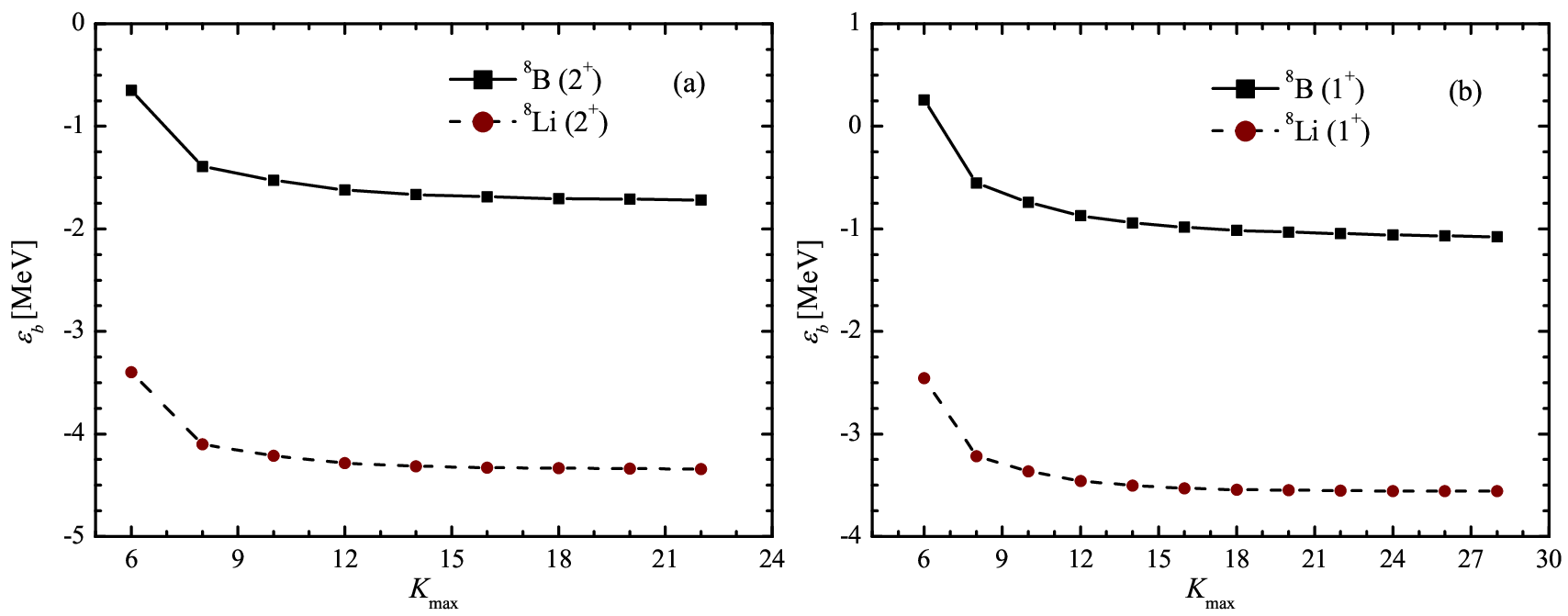} \caption{Convergence of the ground (2$^+$)(a) and first excited (1$^+$) (b) bound state energies
of the $^8{\rm B}$, $^8{\rm Li}$ nuclei with respect to $K_{\rm
max}$.}\label{fig2}
\end{figure}

The three-body bound state wave functions of the $^8{\rm B}$ and $^8{\rm Li}$ nuclei are calculated using the hyperspherical
Lagrange-mesh method \cite{desc03} with the above described $\alpha-N$, $\alpha-^3$He($^3$H), and $p(n)-^3$He($^3$H) two-body
interaction potentials. The parameters values $\hbar^2/2 m_N=20.7343$ MeV fm$^2$, $m_i=A_i m_N$, $A_1=4$, $A_2=3$, $A_3=1$ and
$\hbar$c=197.327 MeV fm are used throughout numerical calculations. The charge numbers are $Z_1=2$, $Z_2=2$, $Z_3=1$ and $Z_1=2$,
$Z_2=1$, $Z_3=0$ for the $^8$B and $^8$Li nuclei, respectively. Actually, the three-body calculations must reproduce the three-body
bound state energies. In our case, the three-body calculations reproduce approximately the experimental energy values in respect to
the $\alpha+^3$He($^3$H)$+p(n)$ three-body threshold, -1.724 MeV and -4.499 MeV  \cite{tilley04} for the $J^{\pi}=(2^+,1)$ ground
states of the $^8{\rm B}$ and $^8{\rm Li}$ nuclei, respectively, without including any three-body forces. The projecting constant
$\Lambda=10^4$ MeV which enables a convergence of the OPP method for eliminating all Pauli forbidden states from the functional space
of the three-body wave functions. In panel (a) of Fig.~\ref{fig2} we present a convergence of the energy values of the ground and
first excited $J^{\pi}=(1^+,1)$ bound states of the $^8\rm{B}$ and $^8\rm{Li}$ nuclei in respect to the maximal hypermomentum
$K_{max}$. The ground state energies are saturated at $K_{max}=22$ for both nuclei. For the 1$^+$ excited bound states, all the
two-body potentials are slightly renormalized by a scaling factor of 1.02 in order to reproduce the experimental energy values of
-0.954 MeV and -3.518 MeV \cite{tilley04}, respectively. This procedure is equivalent to adding a three-body potential and is even
used to describe the ground states of nuclei \cite{desc03}. In panel (b) of Fig.~\ref{fig2} we show a convergence of the energy
values in respect to the $K_{max}$ for the 1$^+$ first excited bound states. One can note here that a convergence of the energy
values for the excited states is slower than for the ground states. The calculated bound state energies are presented in Table
\ref{table4} in comparison with the experimental data ~\cite{tilley04}.
\begin{table}[htbp] \centering
\caption{Calculated bound state energies (MeV) and matter radii (fm).}\label{table4}
\begin{tabular}{cccccccc}
\hline\hline & ~~$^8\rm{B}$~~ & ~~Exp.~~ & ~~$^8\rm{Li}$~~ & ~~Exp.~~ \\
\hline $E$ (2$^+$) & -1.718 & -1.724~\cite{tilley04} & -4.340 & -4.499 ~\cite{tilley04}  \\
$E$ (1$^+$) &-1.075 & -0.954~\cite{tilley04} & -3.559 & -3.518~\cite{tilley04}\\
$<r_{mat}^2>^{1/2}$ & 2.76 & $2.58(6)$ ~\cite{kor18} & 2.56 & $2.50(6)$ ~\cite{alk11} \\
\hline\hline
\end{tabular}
\end{table}

\subsection{Nucleon separation energies of the $^8\rm{B}$ and $^8\rm{Li}$ nuclei}
The proton separation energy of the  $^8\rm{B}$ ground state is a key characteristic for the study of the ANC of the
$^8$B$\rightarrow ^7$Be+$p$ virtual transition. In the $^8$B$=^7$Be+$p$ two-body cluster model the proton separation is easily
reproduced with the help of the high-accuracy Lagrange-mesh and Numerov alghoritms. However, reproducing the proton separation energy
within the three-body model is a rather complex task due to its very small value $E_p$= 0.1375 MeV \cite{tilley04}
and the limited $K_{max}$ in the numerical calculations.

The two-body energies for the ground states of the $^7\rm{Be}$ and $^7\rm{Li}$ nuclei have been calculated using a highly accurate
Lagrange-Laguerre mesh method \cite{baye15} with the same $\alpha-^3$He and $\alpha-^3$H potentials of Ref.~\cite{dub10}, which were
used for the calculation of the three-body energy and wave functions of the $^8\rm{B}$ and $^8\rm{Li}$ nuclei. The calculations have
been done with the number of mesh points $N=40$ and a scaling parameter $h_d=0.40$. The calculated binding energy values of the
ground ($3/2^{-}$) and the first excited ($1/2^{-}$) bound states of the $^7\rm{Be}$ nucleus are 1.5924 MeV and  1.1368 MeV,
respectively. Thus, the theoretical results slightly differ from the experimental data of 1.5866 MeV  and 1.1575 MeV, respectively
\cite{tilley02}. The corresponding binding energies of the ground ($3/2^{-}$) and the first excited ($1/2^{-}$) bound states of the
mirror $^7\rm{Li}$ nucleus are 2.4656 MeV and 1.9943 MeV, close to the experimental data \cite{tilley02}.

Similar to the three-body binding energy, the theoretical proton separation energy $E_p$ of the $^8\rm{B}$ nucleus strongly depends
on the value of the maximal hypermomentum $K_{max}$. The value of $E_p$ is subtracted from the binding energy $E_b(^8{\rm B})$ using
the theoretical estimate $E_b(^7{\rm Be})$=1.5924 MeV and the equation $E_p=E_b(^8{\rm B})- E_b(^7{\rm Be})$. The neutron separation
energy of the $^8$Li  nucleus is estimated in an analogous way: $E_n=E_b(^8{\rm Li})- E_b(^7{\rm Li})$ with $E_b(^7{\rm Li})$=2.4656
MeV in the $\alpha+^3$H channel. In Table \ref{table4a} the calculated values of the three-body binding energies of the $^8\rm{B}$
and $^8\rm{Li}$ nuclei and the corresponding proton and neutron separation energies are presented in dependence on the maximal value
of the hypermomentum $K_{max}$. As can be seen from the table, a good convergence of the results for the proton separation energy is
achieved at $K_{max}$=20 and 22. However, a limitation $K_{max}=12$ \cite{grig98,grig99} does not enable convergent results. In the
case of the $^8\rm{Li}$ nucleus, the neutron separation energy is well reproduced beginning from $K_{max}$=12 due to absence of the
Coulomb interaction between the $^7$Li core and the neutron.

\begin{table}[htbp] \centering
\caption{Dependence of the proton separation energy of the $^8$B g.st. and of the neutron separation energy of the $^8$Li g.st. on
the value of $K_{max}$.} \label{table4a}
\begin{tabular}{ccccccc}
\hline\hline
 $K_{max}$  & 8 & 12 & 16 & 20 & 22 & Exp. \\
\hline
$E_b(^8{\rm B})$ (MeV) & 1.3901  & 1.6221  & 1.6895  & 1.7121  & 1.7175  & 1.7236 \\
$E_p$ (MeV)            & -0.2023 & 0.0297  & 0.0971  & 0.1197  & 0.1251  & 0.1375 \\
$E_b(^8{\rm Li})$ (MeV)& 4.1005  & 4.2846  & 4.3274  & 4.3380  & 4.3399  & 4.4992\\
$E_n$ (MeV)            & 1.6349  & 1.8190  & 1.8618  & 1.8724  & 1.8743  & 2.0323 \\

\hline\hline
\end{tabular}
\end{table}

\subsection{Matter radii}

The root-mean-square (rms) nuclear matter radii $\la r_{mat}^2 \ra ^{1/2}$ of the $^8\rm{B}$ and $^8\rm{Li}$ nuclei can be calculated
with the help of the three-body wave functions. The matter radius of the $^8\rm{B}$ nucleus is defined as:
\begin{eqnarray}
\la r_{mat}^2 \ra = \frac{1}{8} \left( 4 \la r_{\alpha}^2 \ra +3 \la r_{^3\rm{He}}^2 \ra
 +\la r_p^2 \ra +\la \rho^2 \ra  \right), \label{eq13}
\end{eqnarray}
where $\la r_{\alpha}^2 \ra ^{1/2}$= 1.67824(83) fm \cite{nature21}, $\la r_{^3\rm{He}}^2 \ra ^{1/2}$=1.9661(30) fm~\cite{ang13}, and
$\la r_p^2 \ra ^{1/2}$= 0.8406(15) fm \cite{nature26} are experimental charge radii of the corresponding nuclei. The mean square
hyperradius $\la \rho^2 \ra $ is calculated numerically.

\begin{figure}[htbp]
\includegraphics[width=6 cm]{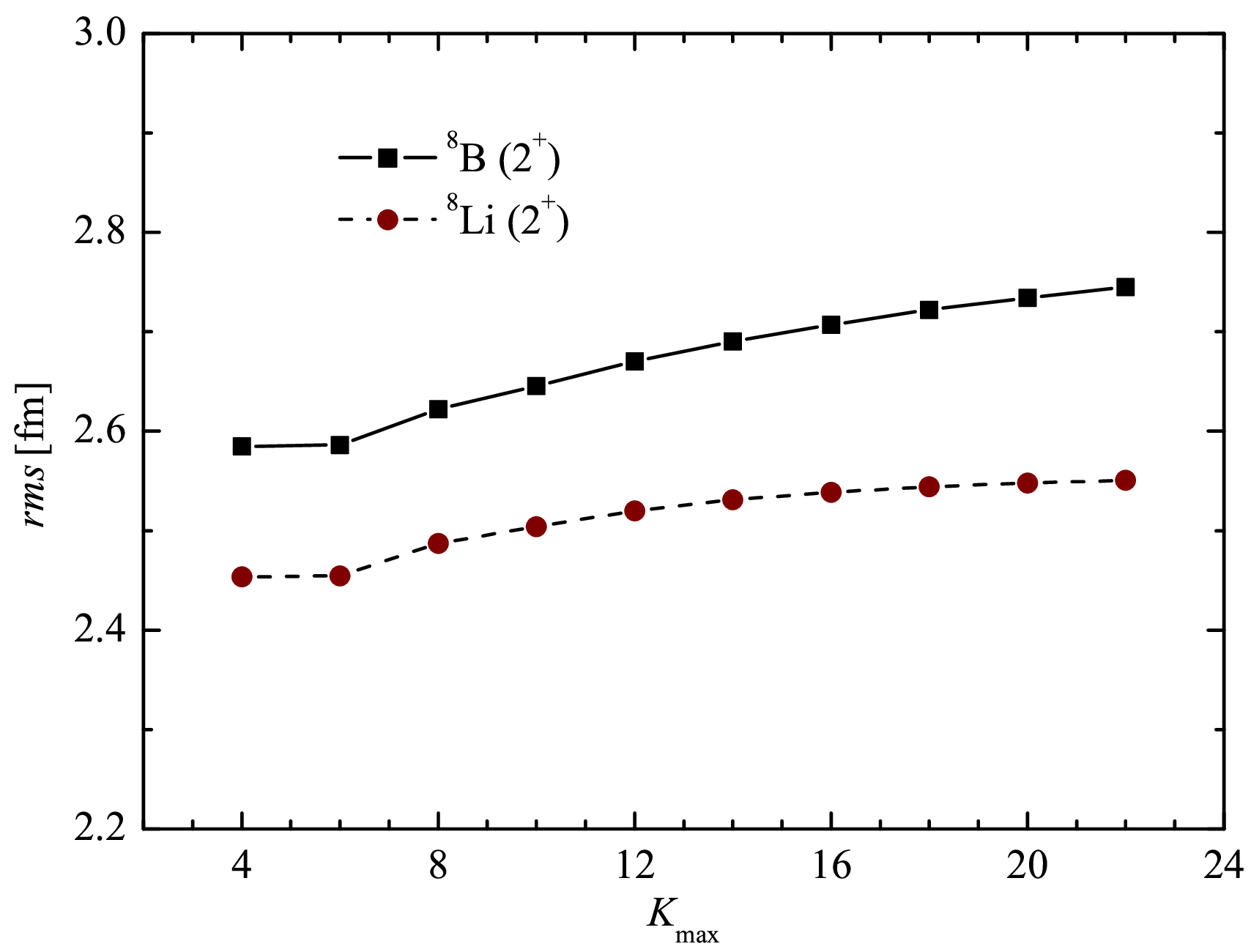}\caption{Convergence of the matter radii of the $^8{\rm B}$ and $^8{\rm Li}$ nuclei
with respect to $K_{\rm max}$.}\label{fig3}
\end{figure}

In the case of the $^8\rm{Li}$ nucleus, the matter radius is calculated using the value $<r_{^3\rm{H}}^2>^{1/2}$=1.7591(363) fm
~\cite{ang13} instead of  $<r_{^3\rm{He}}^2>^{1/2}$ as for the second particle. In Fig.~\ref{fig3} we examine a convergence of the
calculated matter radii for the ground states of the $^8\rm{B}$ and $^8\rm{Li}$ nuclei as a function of $K_{max}$, which varies from
4 up to 22. Within the developed three-body model the matter rms radius of the $^8\rm{B}$ one-proton halo nucleus is overestimated by
about 5$\%$. At the same time, the  rms radius of the $^8\rm{Li}$ nucleus is reproduced within the experimental error bar
~\cite{alk11}. In Table \ref{table4} we present the rms matter radii in comparison with experimental data from
Refs.~\cite{kor18,alk11}. These theoretical results differ slightly from the values 2.56 fm and 2.38 fm obtained using the similar
method in Ref.~\cite{grig98} for the $^8\rm{B}$ and $^8\rm{Li}$ nuclei, respectively. However, one should note that the results of
Ref.~\cite{grig98} have been obtained with the value $K_{max}=12$.

\subsection{ANC and the $S_{17}(0)$ astrophysical factor}
For the estimation of the ANC values within the three-body model we rewrite the $\alpha+^3$He($^3$H)+$N$ three-body bound state wave
function of the $^8$B and $^8$Li nuclei as described in equation (\ref{eq5}), renormalized in relative coordinates $\ve{r}$, $\ve{R}$
in the following form
\begin{multline}
\Psi_3^{JM\pi}(\vec{r},\vec{R},\xi_2,\xi_3) = (\mu_{12}r^2+\mu_{(12)3}R^2)^{-5/4} \\
\times \sum_{\gamma,K} \chi_{\gamma K}^{J\pi}
\left((\mu_{12}r^2+\mu_{(12)3}R^2)^{1/2}\right) \\
\times\left\{{\cal Y}_{l_r l_R}(\hat{r}, \hat{R}) \otimes \chi^S({\xi_2},{\xi_3}) \right\}_{JM} \\
\times \phi^{\ell_r \ell_R}_{K} \left( arctan \left(\frac{\sqrt{\mu_{(12)3}}R} {\sqrt{\mu_{12}}r} \right) \right), \label{eq8}
\end{multline}
where $\rho$ is a hyperradius, $\alpha$ is a hyperangle, $\phi^{\ell_r \ell_R}_{K}(\alpha)$ are the hyperspherical harmonics. The
Jacobi coordinates $\ve{x_3}$ (between the $\alpha$-particle and $^3$He(or $^3$H)) and $\ve{y_3}$ (between the $^7$Be(or $^7$Li) and
the proton (or the neutron)) and the relative coordinates $\ve{r}$, $\ve{R}$ are related as ${\bf x_3}=\sqrt{\mu_{12}}~{\bf r}$ and
${\bf y_3}=\sqrt{\mu_{(12)3}}~{\bf R}$, respectively.

In order to extract the ANC value for the virtual decay $^8{\rm B}\rightarrow^7$Be$+p$ in the $I=1$ or $I=2$ spin fixed channel, the
overlap integral is matched with its asymptotics at large values of the relative distance  $R$ between the $^7$Be core and the
proton:
\begin{multline}  \label{eq16}
I_{I}(R)=\langle \Psi_3(\vec{r},\vec{R},\xi_2,\xi_3) | \psi_I(\vec{r},\hat{R},\xi_2,\xi_3) \rangle_{\vec{r},\hat{R}}
\\ =C_I
W_{\eta;~l_R+1/2}(2\kappa R)/R,
\end{multline}
where $W$ is known Whittaker function, $\kappa$ is the wave number, $\eta$ is the Sommerfeld parameter, and
\begin{multline}
\psi_I(\vec{r},\hat{R},\xi_2,\xi_3)=\frac{u_{2}(r)}{r} \\
\times\left\{\left\{\left\{Y_1(\hat{r}) \otimes \chi_{1/2}(\xi_2) \right\}_{3/2}\otimes \chi_{1/2}(\xi_3)\right\}_I\otimes
Y_{l_R}(\hat{R})\right\}_{2M}
\end{multline}
is an effective wave function. The wave number is calculated at the separation energy 0.1375 MeV of the $^8$B  bound state into the
proton and $^7$Be. In the case of the $^8$Li nucleus, the neutron separation energy is 2.0323 MeV. Integration in Eq.~(\ref{eq16}) is
done over the spatial variable $\vec{r}$, angular variable $\hat R$, and the spin coordinates $\xi_2$ and  $\xi_3$. Since a value of
$l_R=1$ is fixed, we denote the ANC as $C_I$ for simplicity.

\begin{figure}[htbp]
\includegraphics[width=0.45\textwidth]{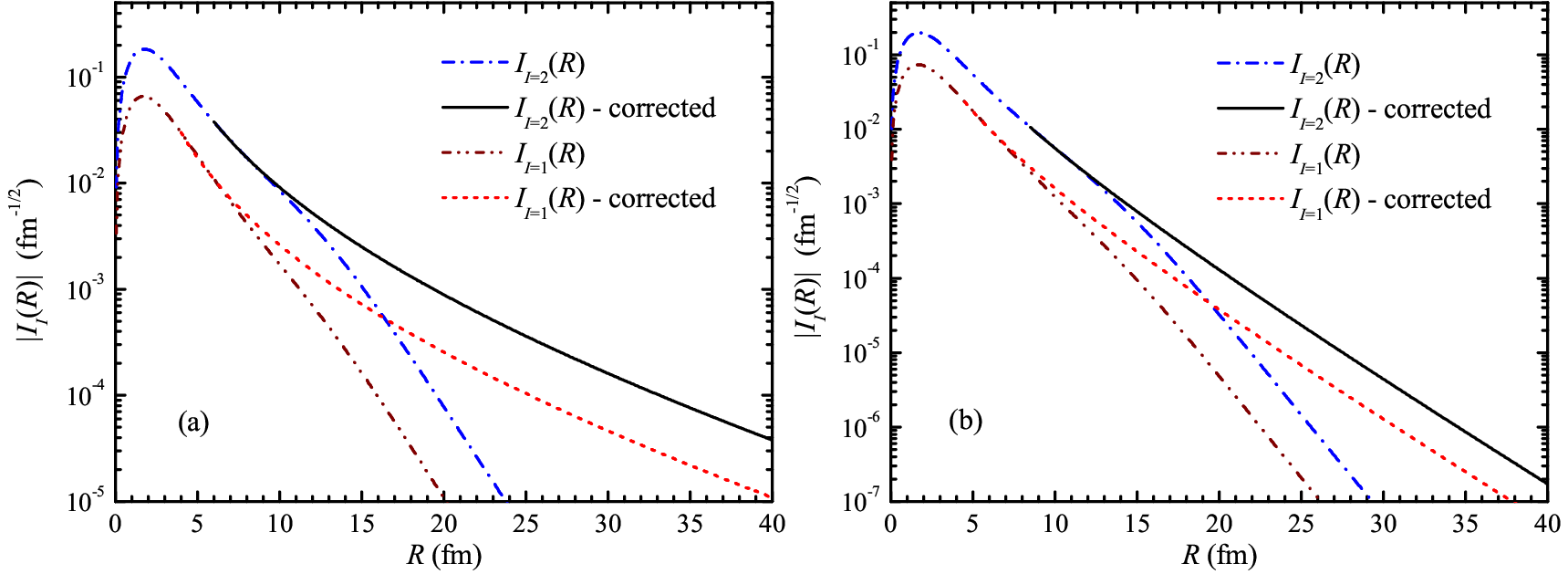} \caption{Overlap integrals in the spin 1 and spin 2 channels with the initial three-body and the corrected
asymptotics for the ground (2$^+$) states of the $^8{\rm B}$ (a) and $^8{\rm Li}$
(b) nuclei.} \label{fig4b}
\end{figure}

In the present three-body model, over the interval 3 - 10 fm $I_I(R)$ follows the expected asymptotic behavior
      $C_I \cdot W_{\eta;~l_R+1/2}(2\kappa R)/R$. In Fig.~\ref{fig4b} the overlap integrals with the initial three-body and the corrected
asymptotics are presented. For the $^8$B(2$^+$) ground state the spin 1 channel asymptotics is corrected at the matching point
$R_0=4$ fm, while for the main spin 2 channel $R_0=6$ fm. In the case of the $^8$Li(2$^+$) ground state the asymptotics is corrected
at the matching points $R_0=4.5$ fm and $R_0=8.5$ fm in the spin 1 and spin 2 channels, respectively.

\begin{figure}[htbp]
\includegraphics[width=0.45\textwidth]{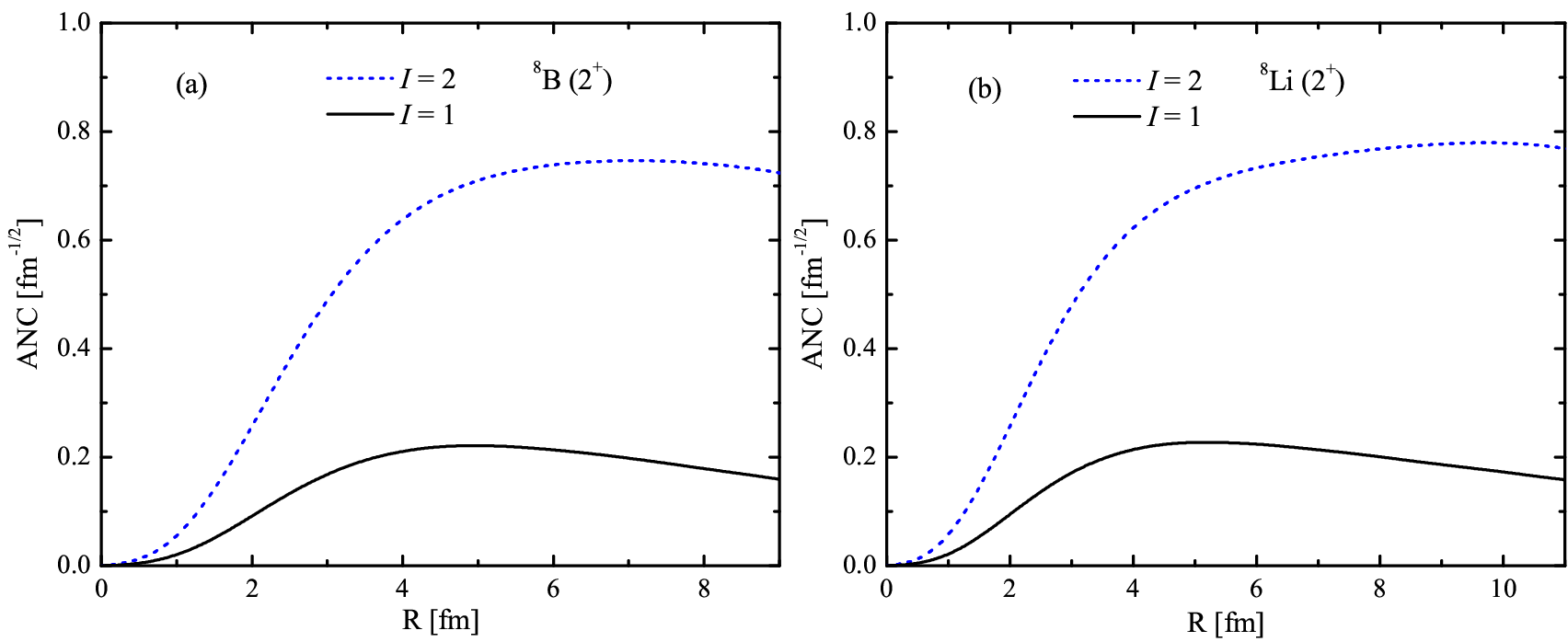} \caption{Calculated ANC values for the ground (2$^+$) states of the $^8{\rm B}$ (a) and $^8{\rm Li}$
(b) nuclei in dependence on the matching point of the overlap integral with its asymptotics.} \label{fig4}
\end{figure}

\begin{table*}[t] \centering
\caption{Values of the squared ANC ($C^2=C^2_{p_{3/2}}+C^2_{p_{1/2}}$) for the $^8$B$\rightarrow ^7$Be+$p$ and astrophysical
$S_{17}(0)$ factor for the $^7$Be(p,$\gamma)^8$B capture reaction from different sources.}\label{table5}
\begin{tabular}{cccccc}
\hline\hline ~~~~~~Method~~~~~~ & ~~~~~~$C^2$, fm$^{-1}$~~~~~~ & ~~~~~~$S_{17}(0)$ eV b~~~~~~ & ~~~~~~Refs.~~~~~~ \\
\hline  Hyperspherical Lagrange-mesh & $0.591\pm0.013$ & $22.492\pm0.503$ &  Present work  \\
\hline Two-body potential model  & $0.538 ^{+0.052}_{-0.050}$ & $20.51^{+2.02}_{-1.85}$ &  \cite{tur21} \\
Microscopic three-  & 0.668 & 25.38 & \cite{desc04}\\
body ($\alpha ^3$He $p$) model  & 0.812 & 30.86 & \cite{desc04}\\

NCSM/CD Bonn 2000 NN & 0.587 & $22.1\pm 1.0$ (at $E$=10 keV) & \cite{nav06}\\
NCSM/RGM & 0.509 & $19.4\pm0.7$ & \cite{nav11}\\
NCSM/ChEFT NN+3N &   & $19.8\pm0.3$ & \cite{nav23}\\

Halo EFT at NLO & $0.564\pm0.023$ & $21.3\pm0.7$ & \cite{zhang15,zhang18}\\
Variational Monte-Carlo {\it ab initio} & $0.537 \pm 0.027$ &    & \cite{nol11} \\

MDWBA $^7$Be$(d,n)^8$B  & $0.613\pm0.060$ & $22.8\pm2.2$ &\cite{olim16} \\

MDWBA $^{10}$B($^7$Be,$^8$B)$^9$Be  &
& &\\
$^{14}$N($^7$Be,$^8$B)$^{13}$C  & $0.465\pm0.041$ & $18.2\pm1.8$ &\cite{tab06}\\

Breakup $^{208}$Pb($^8$B,$p^7$Be)$^{208}$Pb  & 0.548
& $21.7^{+0.37}_{-0.24}{\rm(th)}\pm0.50{\rm(exp)}$ & \cite{oga06} \\

CDCCM $^7$Be$(d,n)^8$B  &
$0.545^{+0.036}_{-0.034}{\rm(th)}\pm0.070{\rm(exp)}$
& $20.96^{+1.4}_{-1.3}{\rm(th)}\pm2.7{\rm(exp)}$ & \cite{oga03} \\

Coulomb breakup A($^8$B,$p^7$Be)A  & $0.450\pm0.072$ & $17.4\pm1.5$ & \cite{tra01} \\

$^{58}$Ni($^8$B,$p^7$Be)$^{58}$Ni  & $0.543\pm0.027$ & $20.8\pm1.1$ & \cite{bel09} \\

R-matrix $^7$Be($p,\gamma)^8$B & $0.518$ & $19.41$ & \cite{huang10} \\
ANC method &  & 17.60 & \cite{nat94} \\
Phenomeno- &
& $20.9\pm0.6{\rm(th)}\pm0.7{\rm(exp)}$  & \cite{jun10} \\

logical &
& $21.2\pm0.7$ & \cite{baby03} \\

way &
& $20.0\pm0.8$ & \cite{buo22} \\

Three-body hyperspherical &
&  &  \\
harmonics method &
& $19.2\pm0.1$ & \cite{grig98} \\

Microscopic shell model &  & 21.4 $\div$ 28.3 & \cite{nat97} \\
GCM &  & 24.5 $\div$ 31.5 & \cite{nat97} \\

Source term approach & 0.389 $\div$ 0.431 &   & \cite{nat10} \\
Updated source term approach & 0.451 &   & \cite{nat13} \\

From solar neutrino fluxes &
& $19.5\pm1.9$ & \cite{tak18} \\

Solar fusion II  &
& $20.8\pm0.7{\rm(th)}\pm1.4{\rm(exp)}$ & \cite{solar2} \\
Solar fusion III  &
& $20.5\pm0.7$ & \cite{solar3} \\
\hline\hline
\end{tabular}
\end{table*}

From above Eq.(\ref{eq16}) one can estimate a value of ANC self-consistently in a theoretical way. At asymptotic distance
$R$ between the $^7$Be ($^7$Li) core and the valence proton (neutron) the ANC value is calculated with the help of the ratio
$C_I=I_{I}(R)\cdot R/W_{\eta;~l_R+1/2}(2\kappa R)$. A physical ANC value is determined from the stable plateau region as a function
 of the matching point $R_0$ ~\cite{kuk95}. In
Fig.~\ref{fig4} we show a dependence of the calculated values of $C_I$  in the spin 1 ($I=1$)  and spin 2 ($I=2$) channels on the
 matching point for the ground states of the $^8$B and $^8$Li nuclei. The stable plateau region in the spin 1 channel of the $^8$B
 nucleus is a range from
3.5 to 7 fm, and the average ANC value $C_{I=1}=0.211 \pm 0.011 $~fm$^{-1/2}$. In the spin 2 channel of this nucleus a stable plateau
begins from 5.0 fm up to 9.0 fm which yields $C_{I=2}=0.739 \pm 0.008$~fm$^{-1/2}$. The corresponding estimates within the {\it ab
initio} NCSM/RGM approach are $C_{I=1}$=0.294~fm$^{-1/2}$ and $C_{I=2}=0.650$~fm$^{-1/2}$, respectively \cite{nav11}. An estimate
0.591 $\pm 0.013$ ~fm$^{-1}$ for the sum of the squared ANC values $C^2=C^2_{I=1}+C^2_{I=2}$ in the present model is consistent with
the NCSM/CD Bonn 2000 NN \cite{nav06}, but larger than the {\it abinitio} NCSM/RGM result of 0.509~fm$^{-1}$ \cite{nav11}.
Additionally, the three-body result is consistent with the result of the halo EFT at NLO within the error bars
\cite{zhang15,zhang18}.

The ANC value is calculated also for the mirror nucleus virtual transition  $^8$Li $->^7$Li+$n$.  In the spin 1 channel, a value
$C_{I=1}=0.220 \pm 0.008$~fm$^{-1/2}$ was obtained in the stable plateau region of (3.5 - 7) fm. And in the spin 2 channel the stable
plateau region (5.0-9.0) fm yields the result $C_{I=2}=0.774\pm 0.008$~fm$^{-1/2}$. For the sum of the squared ANC the present model
yields a value 0.648$\pm 0.013$~fm$^{-1}$, while the microscopic model of Ref.~\cite{desc04} gives 0.747 ~fm$^{-1}$. These numbers
can be compared with the experimental value 0.432$\pm 0.044~$fm$^{-1}$ extracted from the data for the
$^{13}$C($^7$Li,$^8$Li)$^{12}$C neutron transfer reaction~\cite{tra03}, which seems too small.

Now we aim to examine the mirror symmetry of the strong nuclear forces between the two mirror processes $^7$Be(p,$\gamma)^8$B and
$^7$Li(n,$\gamma)^8$Li. The charge symmetry of the strong interaction implies a relation between the asymptotic normalization
coefficients of the mirror nuclei $^8$B and $^8$Li \cite{nat03,nat05,nat06,nat08}. According to Ref.~\cite{nat08} the modified
approximate relation
\begin{equation}
 \mathcal{R}=\left|\frac {C(^8{\rm B})}{C(^8{\rm Li})}\right|^2 \approx \frac{S_p}{S_n} \mathcal{R}_0, \\ \\
 \mathcal{R}_0 = \left|\frac{F_l(ik_p R_N)}{k_p R_N j_l(i k_n R_N)}\right|^2
\end{equation}
is predicted at the nuclear interior radius $R_N=1.3 \cdot 7^{1/3} \approx 2.4868$ fm, where $F_l$ and $j_l$ are regular Coulomb and
spherical Bessel functions, respectively. The parameters $S_p$ and $S_n$ represent the spectroscopic factors for the $^8$B  and
$^8$Li mirror nuclei. The wave numbers $k_p$ and $k_n$ are calculated from corresponding proton and neutron separation energies
$E_p=0.1375$ MeV and $E_n=2.0323$ MeV for the $^8$B and $^8$Li mirror nuclei, respectively. The numerical calculations yield
$S_p/S_n=1.030$  and $\mathcal{R}_0=1.129$. Then the value $\mathcal{R}=\frac{0.591\pm 0.013} {0.648 \pm0.013} \approx 0.912\pm
0.038$ can be compared with the value of $S_p/S_n \cdot \mathcal{R}_0=1.030\cdot 1.129 \approx 1.163$. A difference between the two
numbers 
is understood as a result of breaking of the mirror symmetry of nuclear forces by the three-body dynamical
effects as was explained in Ref.~\cite{nat08} and by the Coulomb interaction.

The astrophysical $S_{17}(0)$ factor is calculated on the basis of the theoretical formalism developed in Ref.~\cite{baye00} which
connects this quantity with the $S$-wave scattering length. This theory is based on the replacement of the initial $S$-wave
scattering and final $P$-wave bound state functions of the $p+^7$Be system by the asymptotic wave functions, the Whittaker and
Coulomb functions, respectively. This approach is consistent with the one proton halo structure of the $^8$B nucleus with a very
small binding energy of 0.1375 MeV. Using the most precise experimental values of the $S$-wave $p+^7{\rm Be}$ scattering lengths from
Ref. \cite{pan19} $a_{0,1}=17.34^{+1.11}_{-1.33}$ fm and $a_{0,2}=-3.18^{+0.55}_{-0.50}$ fm in the spin 1 and spin 2 channels,
respectively, one can estimate the astrophysical $S_{17}(0)$ factor. According to this theory, each term of the sum
$S_{17}(0)=S^{(1)}_{17}(0)+S^{(2)}_{17}(0)$ of the astrophysical factors at zero energy in the spin 1 and spin 2 channels is
expressed as (see Eq.(27) of Ref.~\cite{baye00})
\begin{eqnarray}
S^{(I)}_{17}(0)=38.0 (1-0.0013 a_{0,I}) C_I^2, \ I=1,2, \label{eq15}
\end{eqnarray}
in eV~b with $a_{0,I}$ in fm. In the spin 1 channel with $C_{I=1}=0.211\pm0.011$~fm$^{-1/2}$  the zero-energy $S$-factor
$S^{(1)}_{17}(0)=1.654 \pm 0.173$ {\rm eV~b}. In the spin 2 channel with $C_{I=2}=0.739\pm 0.008$~fm$^{-1/2}$ we obtain
$S^{(1)}_{17}(0)=20.838 \pm 0.472$ {\rm eV~b}. Thus, for the total astrophysical $S(0)$-factor the three-body model yields a value
22.492 $\pm 0.503$ {\rm eV~b}, where the error bars are summed quadratically.

An interesting point is that our theoretical estimate for the $S_{17}(0)$ astrophysical factor has been obtained on the basis of the
asymptotic theory of D. Baye \cite{baye00} using the present ANC values  $C_{I=1}=0.211 \pm 0.011$~fm$^{-1/2}$ and $C_{I=2}=0.739\pm
0.008$~fm$^{-1/2}$, extracted from the three-body wave function, and the values of the $S$-wave scattering lengths in the spin 1 and
spin 2 channels, extracted in Ref.\cite{pan19} within the R-matrix theory. At the same time, in the last work, the R-matrix fit was
performed with the ANC values of 0.315(19) and 0.662(19) in the spin 1 and spin 2 channels from Ref.\cite{nol11}, respectively.
Moreover, a value 0.591 $\pm 0.013$~fm$^{-1}$ of the ANC square in present study is quite different from the estimate of 0.537(27)
\cite{nol11}. The seemed discrepancy can be understood with the help of the statement that the fit to the scattering data was not
highly sensitive to the choice of ANC values in that R-matrix analysis, and although the elastic scattering data was fitted
reasonably well, but the inelastic scattering data could not be explained \cite{pan19}.

The calculated results of the total squared ANC value $C^2$=0.591 $\pm 0.013$~fm$^{-1}$ for the $^8$B$\rightarrow ^7$Be+$p$ virtual
decay and the astrophysical factor $S_{17}$(0)=22.492 $\pm 0.503 {\rm eV~b}$ for the $^7$Be(p,$\gamma)^8$B direct capture reaction
obtained in the present three-body model are compared in Table \ref{table5}  with the results of different theoretical methods and
experimental measurements from
Refs.~\cite{desc04,grig98,nav11,nav06,nav23,zhang15,zhang18,nol11,tur21,olim16,tab06,oga03,oga06,tra01,bel09, huang10,nat94,
jun10,baby03,buo22,nat97,nat10,nat13,tak18,solar2,solar3}.

 As can be seen from the table, the present results are in a good agreement with the results $C^2=0.613\pm0.060$~fm$^{-1}$ and
$S_{17}(0)=22.8\pm2.2$ eV b ~\cite{olim16} extracted from the experimental differential cross-section of the proton transfer reaction
$^7$Be$(d,n)^8$B using the MDWBA approach. The results of the Halo Effective Field Theory \cite{zhang15,zhang18} studies are much
more close to the present three-body model, than the microscopic models of Ref.~\cite{desc04}. Another observation is that the
theoretical studies based on the three-body hyperspherical method \cite{grig98}, R-matrix calculations \cite{huang10}, the ANC method
\cite{nat94}, NCSH/RGM \cite{nav11}, NCSM/ChEFT NN+3N \cite{nav23} yield for the astrophysical $S_{17}(0)$ factor estimates, lower
than 20 ev b, mostly consistent with the Solar Fusion III analysis, while the results of the study of the $^7$Be$(d,n)^8$B transfer
reaction within the CDCCM approach \cite{oga03} are consistent with the present three-body results and with the result of the
indirect method $ S_{17}(0) \approx 19.5 \pm 1.9$ eV b \cite{tak18} extracted from the measured solar $^8$B and $^7$Be neutrino
fluxes from Borexino Collaboration \cite{agos19}. Microscopic shell-model and Generator-coordinate method calculations turn out to be
very sensitive to the description of the NN-interaction \cite{nat97}, while the Source term approach \cite{nat10} and its updated
version \cite{nat13} based on the shell-model wave functions yield quite stable results.

On the other hand, the result of the present three-body model for the astrophysical $S_{17}(0)$ factor is consistent with the Solar
Fusion II recommended value $S_{17}(0)=20.8\pm0.7{\rm(th)}\pm1.4{\rm(exp)}$  eV b, while higher than the recommended value
$S_{17}(0)=20.50\pm0.70$ eV b of the Solar Fusion III analysis ~\cite{solar3}. Surprisingly, the central value of the estimate of the
three-body model is very close to the value 22.4 eV b, used for the most successful new Solar Model BAR2M among other Solar Models,
which predicts both a nearly flat rotation profile in the external part of the radiative region and an increase in the rotation rate
in the solar core \cite{solarmod}. Additionally, the Solar Model BAR2M is consistent with the estimate of the $^8$B neutrino flux by
the Borexino Collaboration \cite{agos19}.

The goal of the three-body model developed in the present work is due-to the use of the most realistic $p-^7$Be scattering data of
S.N.~Paneru et al. \cite{pan19} for the estimation of the $S_{17}(0)$ astrophysical factor within the asymptotic theory of D. Baye
and the self-consistent treatment of the realistic cluster-cluster potentials in the $^8{\rm B}=\alpha+^3{\rm He}+p$ three-body and
$^7{\rm Be}=\alpha+^3{\rm He}$ two-body systems. The asymptotic normalization coefficient was extracted directly from the asymptotics
of the overlap integral between these three-body and two-body wave functions. A consistence of the theoretical result
$S_{17}$(0)=22.492 $\pm 0.503~{\rm eV~b}$ for the $^7$Be(p,$\gamma)^8$B direct capture reaction  with the previous Solar Fusion II
recommended value and with the new most successful Solar Model BAR2M \cite{solarmod} from one hand, and its contradiction to the
Solar Fusion III results indicates, that new experimental efforts are required with an expectation to precisely measure the $^8$B
solar neutrino flux to clarify the situation. Hopefully, the ongoing experimental efforts of the JUNO collaboration will eventually
allow it to perform a model-independent measurement of the $^8$B solar neutrino flux \cite{JUNO}.

\section{Conclusion}
The structure of the $^8$B and $^8$Li nuclei was studied in the framework of the $\alpha+^3$He($^3$H)+$p(n)$ three-body potential
cluster model using the hyperspherical Lagrange-mesh method. The two-body $\alpha-^3$He($^3$H), $\alpha$-nucleon,
$^3$He($^3$H)-nucleon realistic potentials have been used from the literature. A good convergence of the three-body binding energy
and matter radius was demonstrated with the maximal hypermomentum $K_{max}=22$ and $K_{max}=28$ for the ground ($2^+$) and excited
($1^+$) states, correspondingly. The ground states energies have been reproduced without any renormalization of the two-body
potentials or additional three-body potential. For the extraction of the ANC in a self-consistent way, the asymptotics of the overlap
integral of the three-body wave function and the $^7$Be ($^7$Li) core two-body wave function was matched with the Whittaker function.
The ANC values $C_{I=1}=0.211(11)$~fm$^{-1/2}$ and $C_{I=2}=0.739(08)$~fm$^{-1/2}$ are extracted for the virtual decay
$^8$B$\rightarrow ^7$Be+$p$ in the spin 1 and spin 2 channels, respectively. The obtained squared ANC value of 0.591(13) fm$^{-1}$ is
consistent with the results of some analysis of transfer reactions, the NCSM/CD Bonn 2000 NN {\it ab initio} method and halo EFT at
NLO, but larger than the result of the NCSM/RGM {\it ab initio} study.
 For the ANC values of the $^8$Li$\rightarrow ^7$Li+$n$ virtual transition of the mirror
nucleus the estimates $C_{I=1}=0.220(08)$~fm$^{-1/2}$ and $C_{I=2}=0.774(08)$~fm$^{-1/2}$ were obtained.

The ratio of the squared ANC values of the mirror nuclei $^8 {\rm B}$ and $^8{\rm Li}$ estimated within the three-body model implies
a breaking of the mirror symmetry of the strong nuclear forces due to the dynamical three-body effects and the Coulomb interaction
potential.

A detailed three-body analysis of the $^7$Be(p,$\gamma)^8$B direct nuclear capture reaction, a critical process for understanding
solar neutrino fluxes and the Solar Metallicity Problem has been carried out. The zero-energy astrophysical factor
$S_{17}(0)=22.492\pm0.503$ eV b was obtained based on the asymptotic theory of Ref.~\cite{baye00}. It was found that the most
important contribution comes from the spin 2 channel with $S^{(2)}_{17}(0)=20.838 \pm 0.472$, while the spin 1 channel yields
$S^{(1)}_{17}(0)=1.654 \pm 0.173$. The obtained value of the astrophysical $S$ factor at zero energy is in a good agreement with the
Solar Fusion II result $S_{17}(0)=20.8\pm0.7{\rm(th)}\pm1.4{\rm(exp)}$  eV b, but overestimates the recommended value
$S_{17}(0)=20.5\pm0.70$ eV b of the Solar Fusion III ~\cite{solar3}. Fortunately, the result of the developed three-body model is
very consistent with the value 22.4 eV b, used in the most successful new Solar Model BAR2M among different Solar Models
\cite{solarmod}.

\section*{Acknowledgements}
E.M.T. thanks Daniel Baye and Natalia Timofeyuk for useful discussion of presented results.

 \end{document}